\documentclass[conference]{IEEEtran}
\IEEEoverridecommandlockouts

\usepackage{cite}
\usepackage{amsmath,amssymb,amsfonts}
\usepackage{algorithm}
\usepackage{algpseudocode}
\usepackage{amsmath}
\usepackage{graphicx}
\usepackage{textcomp}
\usepackage{xcolor}
\usepackage{balance}
\usepackage[nolist]{acronym}

\def\BibTeX{{\rm B\kern-.05em{\sc i\kern-.025em b}\kern-.08em
    T\kern-.1667em\lower.7ex\hbox{E}\kern-.125emX}}

\begin{document}

\title{Improving Pretrained YAMNet for Enhanced Speech Command Detection via Transfer Learning}
% to add a subscript 1\textsuperscript{st} 1\textsuperscript{st}
\author{\IEEEauthorblockN{Sidahmed Lachenani}
\IEEEauthorblockA{\textit{Research Laboratory in Advanced} \\ \textit{ Electronics Systems LSEA} \\ \textit{ Department of Electrical Engineering} \\
\textit{University of Medea}\\
Medea, Algeria. \\
lachenani.sidahmed@univ-medea.dz}
\and
\IEEEauthorblockN{Hamza Kheddar}
\IEEEauthorblockA{\textit{Research Laboratory in Advanced} \\ \textit{ Electronics Systems LSEA} \\
\textit{ Department of Electrical Engineering} \\
\textit{University of Medea}\\
Medea, Algeria. \\
kheddar.hamza@univ-medea.dz}
\and
\IEEEauthorblockN{Mohamed Ouldzmirli}
\IEEEauthorblockA{\textit{Research Laboratory in Advanced} \\ \textit{ Electronics Systems LSEA} \\
\textit{ Department of Electrical Engineering} \\
\textit{University of Medea}\\
Medea, Algeria. \\
ouldzmirli.mohamed@univ-medea.dz}
}

\makeatletter

\def\ps@headings{%
\def\@oddhead{\parbox[t][\height][t]{\textwidth}{\flushleft

\noindent\makebox[\linewidth]
}
\vspace{0.5cm}
\hfil\hbox{}}%
\def\@oddfoot{\MYfooter}%
\def\@evenfoot{\MYfooter}}
\def\ps@IEEEtitlepagestyle{%
\def\@oddhead{\parbox[t][\height][t]{\textwidth}{
2024 International Conference on Telecommunications and Intelligent Systems (ICTIS)\\

}\hfil\hbox{}}%

\def\@oddfoot{ 979-8-3315-2739-6/24/\$31.00 \textcopyright 2024 IEEE \hfil 
\leftmark\mbox{}}%
\def\@evenfoot{\MYfooter}}

\maketitle

\begin{abstract}

This work addresses the need for enhanced accuracy and efficiency in speech command recognition systems, a critical component for improving user interaction in various smart applications. Leveraging the robust pretrained YAMNet model and transfer learning, this study develops a method that significantly improves speech command recognition. We adapt and train a YAMNet deep learning model to effectively detect and interpret speech commands from audio signals. Using the extensively annotated Speech Commands dataset (speech\_commands\_v0.01), our approach demonstrates the practical application of transfer learning to accurately recognize a predefined set of speech commands. The dataset is meticulously augmented, and features are strategically extracted to boost model performance. As a result, the final model achieved a recognition accuracy of 95.28\%, underscoring the impact of advanced machine learning techniques on speech command recognition. 

This achievement marks substantial progress in audio processing technologies and establishes a new benchmark for future research in the field.

\end{abstract}

\begin{IEEEkeywords}
Speech Command Recognition, Transfer Learning,  YAMNet, Audio Classification
\end{IEEEkeywords}

%correction introduction

%end correction introduction
\section{Introduction}
Speech command recognition is increasingly pivotal in applications such as voice-activated assistants and smart devices, enabling seamless human-computer interaction \cite{devi2023voice}. However, traditional methods often demand significant computational resources and extensive datasets, presenting challenges for real-time performance in resource-limited environments \cite{alam2020survey}.

Recent advances in deep learning (\ac{DL}) have transformed audio processing, making speech recognition models more accurate and resilient \cite{kheddar2024automatic}. Among these is \ac{YAMNet}, a \ac{DL} model optimized for audio classification \cite{ellis2019yamnet}. Pre-trained on extensive audio data, \ac{YAMNet} supports transfer learning (\ac{TL}), enabling adaptation to new tasks with minimal retraining.

This study leverages \ac{TL} to improve speech command recognition by fine-tuning \ac{YAMNet}. Using MATLAB’s Audio Toolbox, we adapted \ac{YAMNet} to recognize a specific set of commands. The Speech Commands dataset \cite{warden2017speech}, a well-regarded benchmark, was used for training and evaluation, providing 32,465 samples across 12 classes.

The goal is to achieve high recognition accuracy by extracting unique acoustic features through advanced \ac{TL} techniques.

This system aims to significantly enhance voice-controlled applications and expand potential use in smart home environments. The findings have implications for advancing robust speech recognition, improving interaction with voice-activated devices, and supporting decision-making in IoT applications.

The contributions of this research include:
\begin{itemize}
    \item \textbf{Enhanced \ac{YAMNet} feature extraction} for speech command recognition, leveraging pre-trained capabilities to improve recognition of specific commands.
    \item \textbf{Comprehensive model configuration testing}, identifying the optimal architecture for effective speech command classification.
    \item \textbf{Targeted hyperparameter optimization}, fine-tuning variables like learning rate and batch size to maximize accuracy on unseen data.
\end{itemize}

The paper is organized as follows: Section \ref{sec2} reviews related studies, establishing the foundation for this research. Section \ref{sec3}-A introduces the dataset, describing composition and preprocessing. Section \ref{sec3}-B details the proposed model, especially \ac{YAMNet} adaptation for improved feature extraction. Section \ref{sec3}-C describes model configuration and performance optimization. Section \ref{sec4} discusses findings and model performance, while Section \ref{sec5} concludes with future research directions and implications for advancing speech recognition.

%enhaced related work
\section{Related work}
\label{sec2}
Recent advancements in speech command recognition have been largely driven by the integration of deep learning (\ac{DL}) techniques. Traditional approaches, such as \acp{HMM} and \acp{GMM}, have been widely employed in speech recognition. However, these methods often fall short in terms of accuracy and efficiency, especially in resource-constrained environments. They typically require substantial amounts of training data and computational resources, which can limit their practicality in real-time applications \cite{chen2022hidden,saravanan2020ensemble}.

The advent of \ac{DL} has revolutionized the field of audio processing, enabling the development of more sophisticated models capable of handling the complexities of speech recognition with greater precision. \Acp{CNN}, \acp{RNN}, and their variants, including \ac{LSTM} networks \cite{hamza2023machine,djeffal2023noise}, have been successfully applied to various speech recognition tasks, including biomedical applications \cite{essaid2024artificial,essaid2025deep}. These models have demonstrated superior performance in capturing the temporal and spectral features of audio signals, resulting in significant improvements in speech recognition accuracy \cite{kheddar2023deep}.

One of the most notable advancements in this domain is the application of \ac{TL}. \ac{TL} involves adapting a pre-trained model, originally trained on a large dataset, to a new but related task with minimal additional training. This approach has proven particularly effective in speech recognition, where large labeled datasets are often scarce. For instance, models like Google’s Speech Commands Model \cite{commands1804dataset} and OpenAI’s Wav2Vec \cite{schneider2019wav2vec} have leveraged \ac{TL} to achieve high performance in recognizing short speech commands with limited training data.

\ac{YAMNet}, developed by Google, is another exemplary \ac{DL} model specifically designed for audio classification tasks. Pre-trained on a large-scale audio dataset, \ac{YAMNet} has been applied in various applications, including environmental sound classification and audio event detection. Its architecture, based on the MobileNetV1 model, is lightweight and well-suited for \ac{TL}, making it an ideal candidate for tasks like speech command recognition in resource-constrained environments.

Recent investigations have demonstrated the effectiveness of \ac{TL} in diverse speech recognition applications. For example, Tang et al. \cite{tang2018deep} illustrate the utilization of deep residual learning for compact keyword detection, highlighting how \ac{TL} optimizes performance with limited datasets. Similarly, McMahan and colleagues \cite{mcmahan2018listening} describe how integrating different datasets, such as UrbanSounds, with speech recognition models enhances noise immunity and robustness. Additionally, in another application, Valliappan \cite{valliappan2024enhancing} uses \ac{YAMNet} to identify gun models from audio clips effectively. This study capitalizes on \ac{YAMNet}'s architecture to extract pivotal features from gunshot sounds, demonstrating its versatility in audio classification for enhancing forensic and security solutions.

This research builds on these advancements by applying \ac{TL} to the \ac{YAMNet} model for the specific task of speech command recognition. It aims to enhance the model’s performance by leveraging the distinct acoustic features of the Speech Commands dataset, contributing to the development of more efficient and accurate voice-controlled systems. Our work not only advances the understanding of \ac{TL} in speech recognition but also provides practical insights into its application in intelligent personal assistants and \ac{IoT} devices.

%end enhaced related work

%New sectio 3
\section{Methodology}
\label{sec3}

\subsection{Dataset Summary}
A notable contribution to the field of speech command recognition is the Speech Commands dataset, introduced by \cite{warden2017speech}. This dataset serves as a standard benchmark for developing and evaluating speech command recognition models. It contains 65,000 one-second utterances of 30 short words, recorded by thousands of individuals, making it particularly suitable for training models to detect simple speech commands in noisy environments.

This study uses a subset of the dataset, including 32,465 audio samples. Each sample is associated with one of 12 categories ("yes", "no", "up", "down", "left", "right", "on", "off", "stop", "go"), along with "unknown" and "background" sounds. To enhance the dataset and improve model robustness, we augmented the audio data by adding background noise segments. This augmentation allows the model to better recognize spoken commands in real-world, noisy environments, thereby increasing both accuracy and reliability in speech command detection tasks.

\begin{figure}[htbp]
\centering 
\includegraphics[height=4.5cm, width=8.5cm]{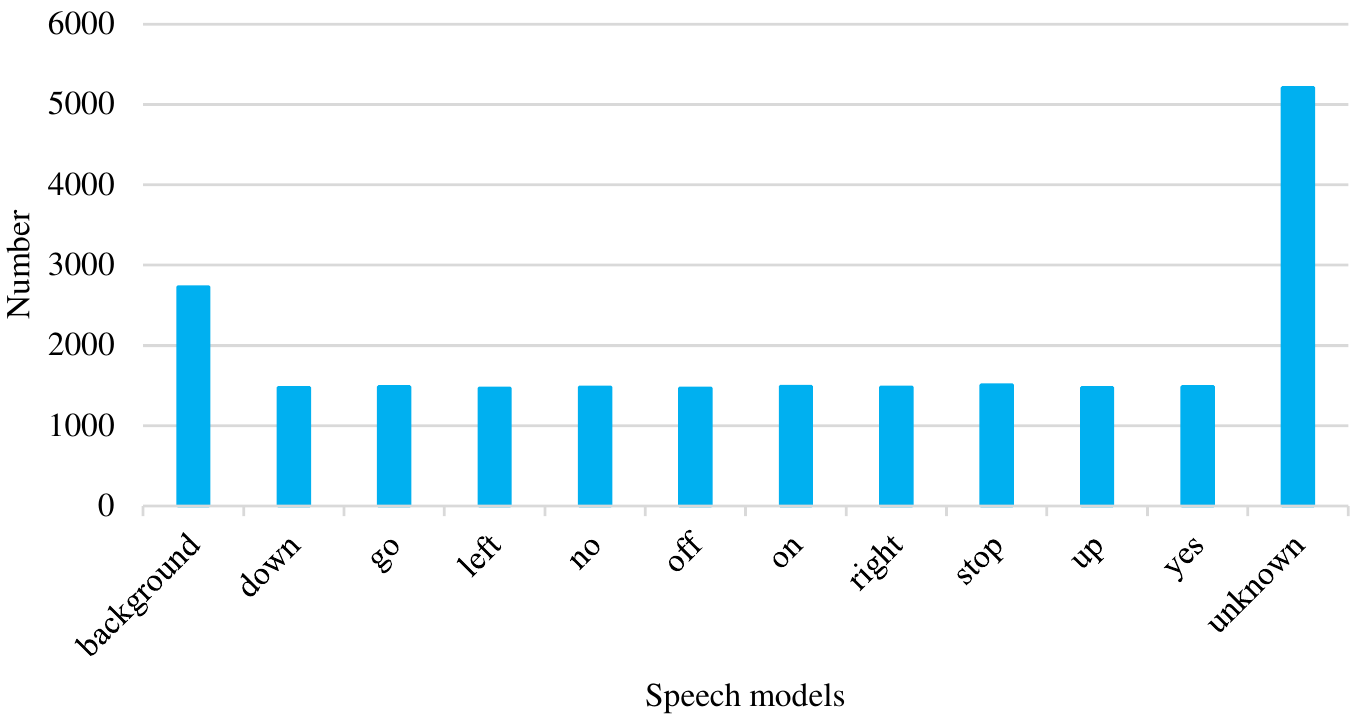}
\caption{Number of audio samples per class.}
\label{fig01}
\end{figure}

The distribution of audio samples across these categories is illustrated in Figure \ref{fig01}, providing a visual overview of the number of samples per category. This dataset is a valuable resource for training and evaluating our simple speech command detection system. By applying \ac{TL}, we leverage \ac{YAMNet} audio classification models to further enhance the accuracy and efficiency of speech command recognition.

\subsection{Data Conditioning}
To ensure compatibility and optimal input for the \ac{YAMNet} audio classification model, data conditioning is performed with a focus on \ac{TL}. The raw speech waveforms are converted to auditory-based spectrograms using a Bark-scale filter bank, which aligns with human auditory perception. Each audio clip is padded and segmented into consistent 1-second intervals.

The signal is resampled to 16 kHz. For a signal \( x(t) \) initially sampled at \( f_s' \), resampling is achieved by:
\begin{equation}
y(t) = x\left( \frac{t}{\alpha} \right)
\end{equation}
where \( \alpha = \frac{f_s'}{f_s} \) and \( f_s = 16000 \, \text{Hz} \).

Following resampling, the signal is divided into frames of 25 ms with a 10 ms hop, producing a number of samples per frame and hop:
\begin{equation}
N_f = T_f \times f_s = 0.025 \times 16000 = 400 \, \text{samples}
\end{equation}
\begin{equation}
N_h = T_h \times f_s = 0.010 \times 16000 = 160 \, \text{samples}
\end{equation}
where \( N_f \) and \( N_h \) are the samples per frame and hop, respectively.

The \ac{STFT} is then applied to each frame to obtain the spectrogram:
\begin{equation}
X(m, k) = \sum_{n=0}^{N_f-1} x[n+mN_h] \cdot w[n] \cdot e^{-j2\pi kn/N_f}
\end{equation}
where \( X(m, k) \) represents the \ac{STFT} at frame \( m \) and frequency bin \( k \), and \( w[n] \) is the Hanning window applied per frame. The spectrogram is calculated by taking the squared magnitude of the \ac{STFT}:
\begin{equation}
S(m, k) = |X(m, k)|^2
\end{equation}
where \( S(m, k) \) denotes the spectrogram value for each frame and frequency bin.

Subsequently, a Bark-scale filter bank transforms the spectrogram into auditory frequency bands, approximating human auditory perception. The Mel frequency \( f_{\text{mel}} \) is given by:
\begin{equation}
f_{\text{mel}} = 2595 \log_{10} \left( 1 + \frac{f}{700} \right)
\end{equation}
where \( f \) represents the frequency in Hz.

To improve generalization, the model balances the representation of background noise and unknown words. Background noise files are segmented into one-second clips to match the duration of speech commands. Additionally, unknown words are included in the dataset to maintain robustness without overwhelming the model with irrelevant data.

This balanced dataset composition allows the model to focus on critical commands, preventing overfitting and enhancing performance in real-world noisy environments where background noise and irrelevant words are prevalent.

\subsection{Presented Model}
\begin{figure}[ht!]
\centerline{\includegraphics[scale=0.6]{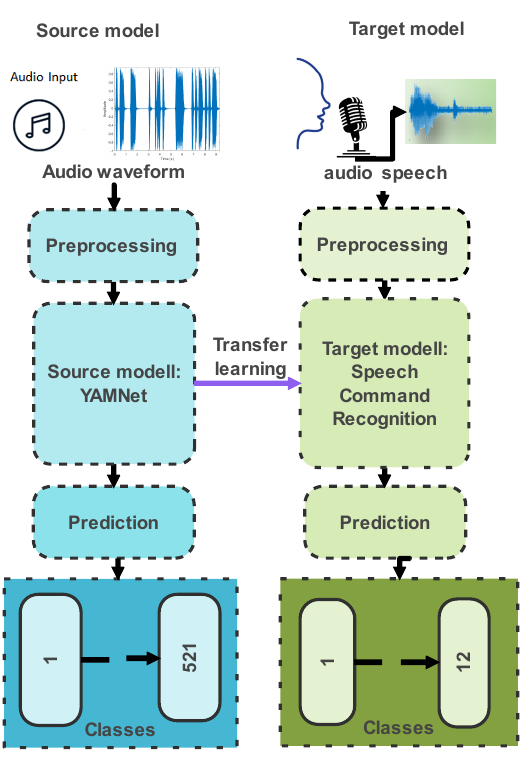}}
\caption{Model Processing Workflow.}
\label{fig02}
\end{figure}

The proposed system for classifying speech commands integrates audio processing and \ac{TL} techniques, as depicted in Figure \ref{fig02}. Initially, audio waveforms are preprocessed into Mel spectrograms, effectively representing audio characteristics. Subsequently, the \ac{YAMNet} model, trained on millions of YouTube videos, acts as a feature extractor. Despite the diversity of its training data, \ac{YAMNet}'s comprehensive learning from such a vast dataset enhances its ability to generalize from limited data, making it particularly effective when applied to smaller datasets through \ac{TL}. MATLAB software then constructs a series of fully connected layers, including a custom layer tailored to \ac{YAMNet}'s outputs. These layers are responsible for classifying Speech Commands based on the extracted features.

\begin{algorithm}
\caption{Speech command classification using \ac{YAMNet} and neural network} \label{alg:algorithm1}
\begin{algorithmic}[1]
\State \textbf{Audio Dataset Preprocessing:}
\State Load audio samples with associated labels.
\State Convert samples to mono and resample to 16 kHz.
\State Augment the dataset by adding background noise segments to balance training.

\State \textbf{Data conditioning:}
\State Convert each audio sample to a Mel-spectrogram with the following parameters:
\begin{itemize}
    \item \textit{Segment Duration:} 1 second
    \item \textit{Frame Duration:} 25 ms
    \item \textit{Hop Duration:} 10 ms
    \item \textit{Number of Frequency Bands:} 50
\end{itemize}
\State Extract high-level features (Spectrogram).

\State \textbf{Constructing the \ac{YAMNet} Neural Network Model:}
\State Construct the neural network architecture using a combination of convolutional layers, batch normalization, and ReLU activation:
\begin{itemize}
    \item \textit{Input Layer:} Takes the extracted feature dimensions as input.
    \item \textit{Hidden Layers:} Multiple convolutional layers with batch normalization and activation.
    \item \textit{Output Layer:} A fully connected layer with softmax activation for classification into speech command categories.
\end{itemize}
\State Modify the output classes from 521 to 12 classes.
\State \textbf{Dataset Segmentation and Model Training:}
\State Segment the dataset into 80\% training and 20\% validation sets.
\State Compile the model using the Adam optimizer and a mini-batch size of 128.

\State \textbf{Model Compilation and Training:}
\State Train the neural network on the extracted embeddings.
\State Assess and validate the model accuracy using the validation embeddings.

\State \textbf{Evaluation and Validation:}
\State Evaluate the model performance using the validation dataset.
\State Calculate the training and validation errors.
\end{algorithmic}
\end{algorithm}

By employing \ac{YAMNet} as a feature extractor and using \ac{TL}, a novel technique for categorizing speech commands is proposed. The dataset includes between 1,471 and 5,208 samples for each speech command type to ensure robustness and prevent overfitting during the training and validation phases. This process maintains a balance between early layers that learn low-level features and later layers that are more task-specific. Hold-out validation is employed, splitting the dataset into a training set and a validation set. The training set is used to refine the model, while the validation set assesses its performance. This method ensures a balanced approach to training machine learning models, typically splitting data 80\% for training and 20\% for validation to optimize accuracy.

The model is trained for 15 epochs using the Adam optimizer, with an initial learning rate of 0.0003. Training was conducted with a mini-batch size of 128 and included a validation dataset for performance assessment. The training options were set to shuffle the data every epoch and provided visual feedback on the training progress. The validation frequency was automatically set to occur once every epoch, calculated based on the number of training samples.

For more details, and as shown in Algorithm~\ref{alg:algorithm1}, the proposed approach for speech command classification begins with audio preprocessing, converting raw audio samples into Mel spectrograms. The \ac{YAMNet} model is then applied as a feature extractor, and the extracted features are used to train a neural network for speech command recognition.

\subsection{Metrics Used}

In this study, we employed several important metrics to evaluate the performance of our speech command recognition system. These metrics provide a comprehensive framework for assessing the model’s ability to accurately classify speech commands while minimizing errors. Table \ref{tab0} summarizes the key metrics used, along with their descriptions and importance in evaluating the system \cite{kheddar2024deep,gueriani2024enhancing}

%new metrics
\begin{table}[htbp]
\scriptsize
\caption{Summary of Evaluation Metrics for Model Performance.}
\label{tab0}
\centering
\begin{tabular}{|p{1cm}|p{2cm}|p{4.5cm}|}
\hline
\textbf{Metric} & \textbf{Formula} & \textbf{Description} \\ \hline
Accuracy & $\frac{\text{TP+TN}}{\text{TP+TN+FP+FN}}$ & Correct classifications among all categories, indicating overall effectiveness. \\ \hline
Precision & $\frac{\text{TP}}{\text{TP+FP}}$ & Ratio of true positives to all predicted positives, lowering false positives. \\ \hline
Recall & $\frac{\text{TP}}{\text{TP+FN}}$ & True positives out of actual positives, reducing false negatives. \\ \hline
F1 Score & $2 \times \frac{\text{Precision} \times \text{Recall}}{\text{Precision + Recall}}$ & Harmonic mean of precision and recall, balancing both metrics. \\ \hline
Specificity & $\frac{\text{TN}}{\text{TN+FP}}$ & True negatives out of actual negatives, minimizing false positives. \\ \hline
\end{tabular}
\end{table}

%These metrics together provide a thorough evaluation of the model’s performance from multiple angles, ensuring that the speech command recognition system is both accurate and reliable in diverse real-world applications.

%end new metrics

%New Section 4
\section{Result}
\label{sec4}

In this study, \ac{YAMNet}, an advanced audio command classification model, was utilized to enhance speech command detection. A total of 22,770 audio samples, divided into 12 distinct classes, were employed for the research, all sourced from the Speech Commands Dataset \cite{warden2017speech}. Each audio sample was transformed into a Mel spectrogram, serving as input for \ac{YAMNet}'s feature extraction process, allowing for precise and optimized audio data analysis.

\subsection{Experimental Setup}
To further improve the model’s performance, multiple trials were conducted to avoid overfitting and restore optimal weights. After proper data augmentation, the dataset was split into training and validation sets with an 80-20\% ratio, ensuring better speech command recognition in real-world scenarios. The validation was conducted on a system equipped with an NVIDIA GeForce GT 710 GPU, using MATLAB R2022b software. All experiments were run on a computer with an Intel Core i5-7500 CPU at 3.40 GHz and 8 GB of RAM, with the GPU handling all computational tasks.

\subsection{Training Configuration}
The established framework of a MATLAB example was utilized as a benchmark for developing an advanced speech command recognition system~\cite{mathworks2024deeplearn}. By integrating \ac{YAMNet} with MATLAB’s \ac{DL} Toolbox, \ac{TL} techniques were leveraged to fine-tune the model specifically for audio classification tasks. The MATLAB example provided a structured approach to training using the Adam optimizer, a mini-batch size of 128, and a tailored learning rate of 0.0003 across 15 epochs. This setup allowed for a systematic enhancement of the model’s capability to recognize a variety of speech commands from the Google Speech Commands Dataset, setting a high standard and benchmark for accuracy and efficiency in ongoing research endeavors.

\subsection{Model Performance Overview}
In addition to the previously presented results, the efficiency of our classification model was evaluated using several metrics. Table \ref{tab1} summarizes the performance in terms of accuracy, F1 score, and precision score for each loss function. As shown in the table, notable variations were observed depending on different configurations. The analysis in the \ac{TL} section provided particularly insightful findings. Starting with an accuracy of 20\%, precision increased to 90\% as the number of epochs approached 1, and the highest accuracy of 95.28\% was reached after 15 epochs. These results suggest that increasing the number of epochs initially improves accuracy.

\subsection{Training Progress Analysis}
\begin{figure}[htbp]
\centerline{\includegraphics[scale=0.9]{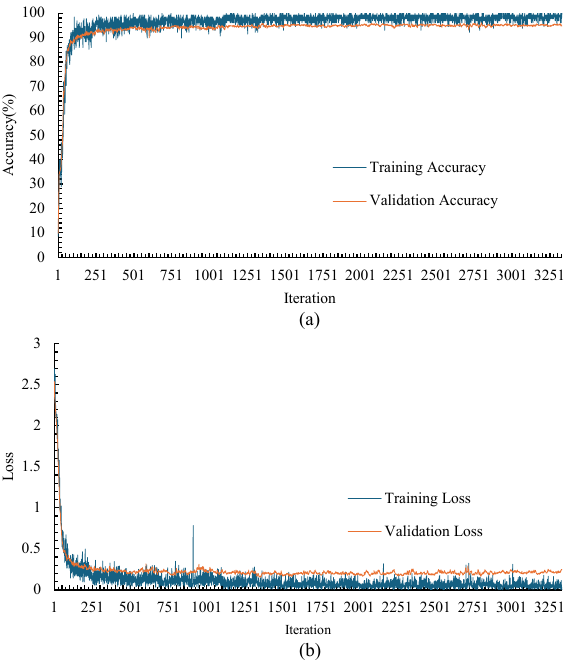}}
\caption{YAMNet training and validation performance metrics progress: (a) Accuracy , (b) Loss.}
\label{fig03}
\end{figure}

The training progress of \ac{YAMNet}, as shown in Figure \ref{fig03}, illustrates both the accuracy and loss curves over the training iterations. The top plot shows the accuracy progression for the training and validation sets, while the bottom plot displays the corresponding loss values.

In the accuracy plot, we observe a steep increase during the initial phase of training, with the model quickly improving performance and reaching approximately 90\% accuracy within the first few hundred iterations. As training progresses, accuracy continues to increase gradually, stabilizing around 95\% for both training and validation data. This consistency between the training and validation accuracy indicates that the model is not overfitting and generalizes well to unseen data.

In the loss plot, a similar trend can be observed. Initially, the training loss decreases rapidly, suggesting effective learning and classification error minimization. The validation loss follows a similar pattern, converging at a low value, which further confirms the model’s ability to perform well on the validation set.

Notably, the smoothness of the training curves indicates the stability of the learning process, likely due to well-tuned hyperparameters such as learning rate and batch size. The close alignment of the training and validation curves throughout the process strongly indicates that the model has generalized well without overfitting, achieving high performance in recognizing speech commands.
\subsection{In-Depth Analysis of Key Metrics}

Our model's performance is evaluated using key metrics: accuracy, precision, recall, F1 score, and specificity, each offering insight into its strengths in speech command classification.

\textbf{Accuracy:} Achieving 95.28\% accuracy demonstrates the model's reliability across all classes, confirming its ability to capture essential features for correct classification. This underscores the value of transfer learning with YAMNet’s pretrained architecture.

\textbf{Precision:} With a precision of 95.08\%, the model effectively minimizes false positives, which is critical in voice-activated applications where incorrect activations could disrupt user experience.

\textbf{Recall:} A recall of 94.43\% indicates the model’s capability to identify nearly all command instances, essential for ensuring valid commands are not missed in real-world applications.

\textbf{F1 Score:} Reaching 94.57\%, the F1 score shows a balanced performance between precision and recall, making the model highly effective in applications requiring both sensitivity and accuracy.

\textbf{Specificity:} At 99.49\%, high specificity demonstrates the model’s robustness in handling background noise and non-command audio, essential for diverse real-world environments.

These metrics confirm the model’s strong predictive power and reliability in accurately categorizing audio commands.

\begin{table}[htbp]
\caption{Performance Metrics of Audio Classification Models. \label{tab1}}
\centering
\setlength{\tabcolsep}{4pt} % Adjust column spacing
\renewcommand{\arraystretch}{1.2} % Adjust row spacing
\begin{tabular}{|c|c|c|c|c|c|}
\hline
\ & \textbf{Accuracy} & \textbf{Precision} & \textbf{Recall} & \textbf{F1 Score} & \textbf{Specificity} \\ \hline
\ac{YAMNet}         & 0.9528           & 0.9508             & 0.9443         & 0.9457           & 0.9949             \\ \hline
\ac{DL} \cite{mathworks2024deeplearn}  & 0.9441           & 0.9446             & 0.9441         &   0.9441           & 0.9949              \\ \hline
\end{tabular}
\end{table}

\subsection{Confusion Matrix Analysis}
The classification performance of our model is thoroughly captured in the confusion matrix, which illustrates the accuracy for each speech command category. Noteworthy performances are seen in commands like "right" and "yes," with high accuracies of 97.27\% and 96.17\%, respectively, demonstrating the model’s strength in recognizing these commands. Conversely, categories such as "go," "no," and "on" display lower accuracy rates of 89.23\%, 92.22\%, and 89.11\%, respectively, suggesting potential areas for refinement. The confusion matrix is instrumental in identifying both the strengths and weaknesses of the model, providing a roadmap for targeted improvements in future iterations.

\begin{figure}[htbp]
\centerline{\includegraphics[scale=0.35]{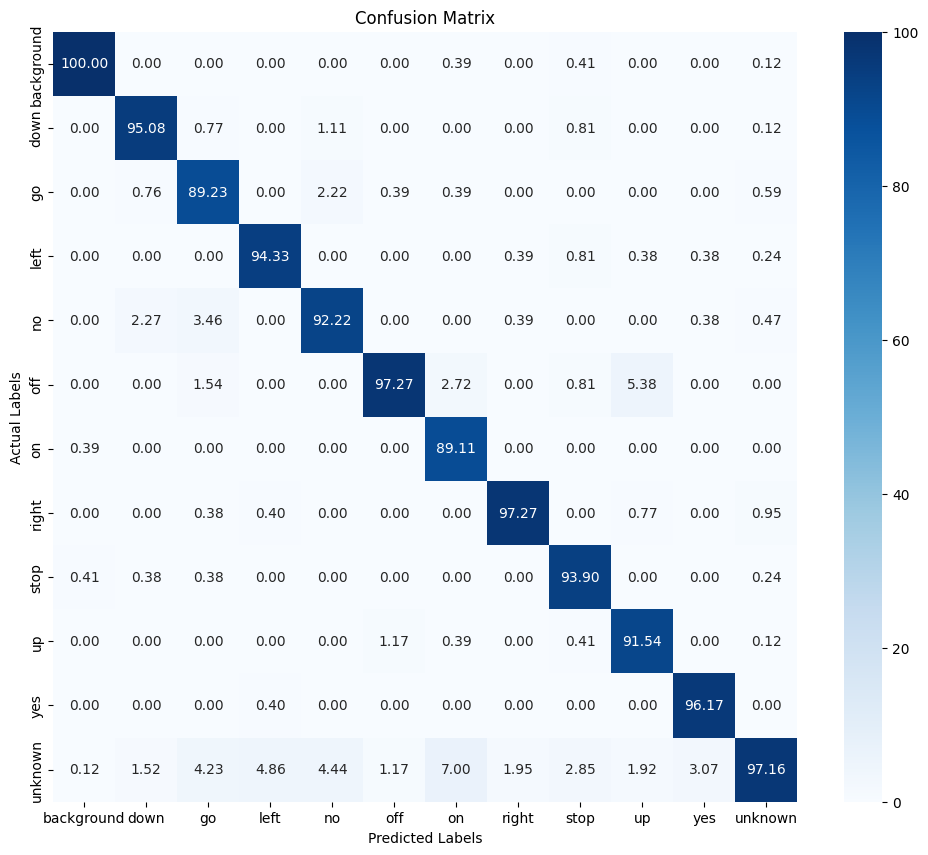}}
\caption{Model’s confusion matrix for speech command.}
\label{fig04}
\end{figure}

As shown in Figure \ref{fig04}, the confusion matrix offers a clear visualization of the classification performance across the 12 command categories. Each value represents the percentage of times a predicted class matches the true instance of another class, with a high diagonal percentage indicating strong classification performance for specific commands.

While commands such as "background" (100\%), "right" (97.27\%), and "yes" (96.17\%) demonstrate high accuracy, other classes like "go" and "on" experience more frequent misclassifications, particularly confusing with "no" and "off." This suggests that acoustic similarities between certain commands may cause confusion, highlighting areas where feature refinement or additional dataset augmentation could improve differentiation.

The strong performance of the "unknown" class (97.16\%) and "background" noise (100\%) also underscores the model’s robustness in rejecting irrelevant or noisy inputs, ensuring reliability in real-world environments where non-command audio is common.

\section{Conclusion and Future Directions}
\label{sec5}

This study presents a speech command recognition system developed using transfer learning with the \ac{YAMNet} audio classification model. Our approach demonstrated strong performance in recognizing predefined speech commands, achieving high accuracy at 95.28\%, an F1 score of 94.57\%, and a precision of 95.08\%. By fine-tuning the \ac{YAMNet} model and strategically modifying its architecture, we improved the model’s ability to capture complex patterns in audio signals, underscoring the effectiveness of transfer learning for speech command recognition, particularly in resource-constrained environments.

Future work will focus on further enhancing YAMNet’s performance by adding layers to support deeper audio representation learning and by fine-tuning hyperparameters to improve training convergence and robustness to variations in audio signals. Additionally, we plan to incorporate larger models—such as Transformers \cite{djeffal2023automatic}, \ac{LLM} \cite{kheddar2024transformers}, and more diverse datasets—to increase the model’s generalization capabilities. Training on more varied data will enable the model to be more resilient to different accents, background noises, and recording conditions, improving its applicability in real-world scenarios. Investigating adversarial attacks \cite{noureddine2023adversarial} and intensive data hiding in speech parameters \cite{kheddar2022speech} is crucial, as they significantly reduce command detection performance.

\section*{Acknowledgment}

This work is supported by the Directorate General for Scientific Research and Technological Development (DGRSDT) of Algeria.

\balance
\bibliographystyle{elsarticle-num}
\bibliography{bibliography}% common bib file

\begin{thebibliography}{10}
\expandafter\ifx\csname url\endcsname\relax
  \def\url#1{\texttt{#1}}\fi
\expandafter\ifx\csname urlprefix\endcsname\relax\def\urlprefix{URL }\fi
\expandafter\ifx\csname href\endcsname\relax
  \def\href#1#2{#2} \def\path#1{#1}\fi

\bibitem{devi2023voice}
M.~Devi, K.~M. Shahriar, I.~Ko, Voice recognition technologies: Comparative analysis and potential challenges in future implementation, Internet E-commerce Research 23~(6) (2023) 285--308.

\bibitem{alam2020survey}
M.~Alam, M.~D. Samad, L.~Vidyaratne, A.~Glandon, K.~M. Iftekharuddin, Survey on deep neural networks in speech and vision systems, Neurocomputing 417 (2020) 302--321.

\bibitem{kheddar2024automatic}
H.~Kheddar, M.~Hemis, Y.~Himeur, Automatic speech recognition using advanced deep learning approaches: A survey, Information Fusion (2024) 102422.

\bibitem{ellis2019yamnet}
D.~Ellis, Yamnet: A pretrained audio event classifier (2019).

\bibitem{warden2017speech}
P.~Warden, Speech commands: A public dataset for single-word speech recognition, Dataset available from http://download. tensorflow. org/data/speech\_commands\_v0 1 (2017).

\bibitem{chen2022hidden}
Y.~Chen, A hidden markov optimization model for processing and recognition of english speech feature signals, Journal of Intelligent Systems 31~(1) (2022) 716--725.

\bibitem{saravanan2020ensemble}
P.~Saravanan, E.~Sri~Ram, S.~Jangiti, E.~Ponmani, L.~Ravi, V.~Subramaniyaswamy, Ensemble gaussian mixture model-based special voice command cognitive computing intelligent system, Journal of Intelligent \& Fuzzy Systems 39~(6) (2020) 8181--8189.

\bibitem{hamza2023machine}
A.~Hamza, D.~Addou, H.~Kheddar, Machine learning approaches for automated detection and classification of dysarthria severity, in: 2023 2nd International Conference on Electronics, Energy and Measurement (IC2EM), Vol.~1, IEEE, 2023, pp. 1--6.

\bibitem{djeffal2023noise}
N.~Djeffal, D.~Addou, H.~Kheddar, S.~A. Selouani, Noise-robust speech recognition: A comparative analysis of lstm and cnn approaches, in: 2023 2nd International Conference on Electronics, Energy and Measurement (IC2EM), Vol.~1, IEEE, 2023, pp. 1--6.

\bibitem{essaid2024artificial}
B.~Essaid, H.~Kheddar, N.~Batel, M.~E. Chowdhury, A.~Lakas, Artificial intelligence for cochlear implants: Review of strategies, challenges, and perspectives, IEEE Access (2024).

\bibitem{essaid2025deep}
B.~Essaid, H.~Kheddar, N.~Batel, M.~E. Chowdhury, Deep learning-based coding strategy for improved cochlear implant speech perception in noisy environments, IEEE Access (2025).

\bibitem{kheddar2023deep}
H.~Kheddar, Y.~Himeur, S.~Al-Maadeed, A.~Amira, F.~Bensaali, Deep transfer learning for automatic speech recognition: Towards better generalization, Knowledge-Based Systems 277 (2023) 110851.

\bibitem{commands1804dataset}
S.~Commands, A dataset for limited-vocabulary speech recognition, URL: https://arxiv. org/abs/1804.03209 (28.12. 2020) (1804).

\bibitem{schneider2019wav2vec}
S.~Schneider, A.~Baevski, R.~Collobert, M.~Auli, wav2vec: Unsupervised pre-training for speech recognition, arXiv preprint arXiv:1904.05862 (2019).

\bibitem{tang2018deep}
R.~Tang, J.~Lin, Deep residual learning for small-footprint keyword spotting, in: 2018 IEEE International Conference on Acoustics, Speech and Signal Processing (ICASSP), IEEE, 2018, pp. 5484--5488.

\bibitem{mcmahan2018listening}
B.~McMahan, D.~Rao, Listening to the world improves speech command recognition, in: Proceedings of the AAAI Conference on Artificial Intelligence, Vol.~32, 2018.

\bibitem{valliappan2024enhancing}
N.~H. Valliappan, S.~D. Pande, S.~R. Vinta, Enhancing gun detection with transfer learning and yamnet audio classification, IEEE Access (2024).

\bibitem{kheddar2024deep}
H.~Kheddar, M.~Hemis, Y.~Himeur, D.~Meg{\'\i}as, A.~Amira, Deep learning for steganalysis of diverse data types: A review of methods, taxonomy, challenges and future directions, Neurocomputing (2024) 127528.

\bibitem{gueriani2024enhancing}
A.~Gueriani, H.~Kheddar, A.~C. Mazari, Enhancing iot security with cnn and lstm-based intrusion detection systems, in: 2024 6th International Conference on Pattern Analysis and Intelligent Systems (PAIS), IEEE, 2024, pp. 1--7.

\bibitem{mathworks2024deeplearn}
{The MathWorks Inc.}, \href{https://ch.mathworks.com/help/deeplearning/ug/ deep-learning-speech-recognition.html}{Deep learning for speech recognition} (2024).
\newline\urlprefix\url{https://ch.mathworks.com/help/deeplearning/ug/ deep-learning-speech-recognition.html}

\bibitem{djeffal2023automatic}
N.~Djeffal, H.~Kheddar, D.~Addou, A.~C. Mazari, Y.~Himeur, Automatic speech recognition with bert and ctc transformers: A review, in: 2023 2nd International Conference on Electronics, Energy and Measurement (IC2EM), Vol.~1, IEEE, 2023, pp. 1--8.

\bibitem{kheddar2024transformers}
H.~Kheddar, Transformers and large language models for efficient intrusion detection systems: A comprehensive survey, arXiv preprint arXiv:2408.07583 (2024).

\bibitem{noureddine2023adversarial}
K.~Noureddine, H.~Kheddar, M.~Maazouz, Adversarial example detection techniques in speech recognition systems: A review, in: 2023 2nd International Conference on Electronics, Energy and Measurement (IC2EM), Vol.~1, IEEE, 2023, pp. 1--7.

\bibitem{kheddar2022speech}
H.~Kheddar, A.~C. Mazari, G.~H. Ilk, Speech steganography based on double approximation of lsfs parameters in amr coding, in: 2022 7th International Conference on Image and Signal Processing and their Applications (ISPA), IEEE, 2022, pp. 1--8.

\end{thebibliography}

\begin{acronym}[FedNST]
\acro{DL}{deep learning}
\acro{TL}{transfer learning}
\acro{YAMNet}{youtube audio model network}
\acro{IoT}{Internet of Things}
\acro{CNN}{convolutional neural network}
\acro{RNN}{recurrent neural network}
\acro{LSTM}{long short-term memory}
\acro{HMM}{hidden Markov model}
\acro{GMM}{Gaussian mixture model}
\acro{STFT}{short-time Fourier transform}
\acro{IoT}{internet of things}
\acro{LLM}{large language model}
\end{acronym}

\end{document}